# Modeling of the long-time asymptotic dynamics of a point-like object

Marijan Ribarič[1] and Luka Šušteršič[2]

Jožef Stefan Institute, p.p. 3000, 1001 Ljubljana, Slovenia

**Abstract.** We introduce four original concepts.  First, the point-like object (PO) specified as a classical extended real object whose response to an external force is aptly specified solely by the trajectory of a single point, whose velocity eventually stops changing after the cessation of the external force.  Second, the dynamic models of an PO  that generalize Newton's second law by the explicit modeling of PO-acceleration by nonlinear functions of the external force.  Third, the long-time asymptotic dynamics of an PO (LTAD)  modeled by polynomials in time-derivatives of the external force (by LTAD-models). To make LTAD-models we do not need to know the PO equation of motion. Given the PO equation of motion, without solving it, we can calculate the appropriate LTAD-models, but not vice versa.
Fourth, the asymptotic differential equations about the LTAD.  They are equivalent to the  LTAD-models, but not to the PO equation of motion.

To resolve the conceptual controversy surrounding the relativistic Lorentz-Abraham Dirac equation, we interpret this equation as an asymptotic differential equation about the LTAD  of an electrified PO, and not as a differential equation of motion for an electrified PO.

---

[1] Corresponding author.  Tel.: +386 1 256 3336. E-mail address: marjan.ribaric@masicom.net.

[2] E-mail address: luka.sustersic@ijs.si.

Modeling of the long-time asymptotic dynamics of a point-like object

1, **Introduction**

We consider the mathematical modeling of dynamics of particular classical *extended* real objects, which we name "point-like objects" (POs). We specify an PO as follows:

a) A PO is a classical real object whose response to an external force is aptly specified solely by the trajectory of a single point, which we name "the PO-position".

b) The acceleration of the PO-position is determined by the external force acting on it.

c) The PO-velocity eventually stops changing after the cessation of the external force.

So far the dynamics of an PO has been modeled by so-named "point-mass", whose acceleration is specified by Newton's second law of motion, dividing the force acting on the point-mass by its mass, its sole kinetic constant. However, an PO may be a very complicated real object such as a planet, train, ship, half-empty bottle, chain, particle . . . , thus we may oversimplify the situation by using a point-mass to model its dynamics. So we address partly this question by introducing an innovative mathematical framework for modeling of the long-time asymptotic dynamics of such POs as $t \nearrow \infty$ , e.g. of a cyclically moving PO, by polynomials in time-derivatives of the external force. Note that *the long-time asymptotic of dynamics at a cyclic PO-movement is the very same dynamics* !

There are many real systems consisting of POs , each of which is treated as the point-mass by taking into account only its mass, e.g. in astronomy, and in classical mechanics. Were we to take some account also of the supplementary kinetic properties of each PO, we might get a better dynamic model of such a system. Thus we introduce such dynamic models of POs that generalize Newton's second law by explicitly specifying the acceleration of an PO as a possibly nonlinear function of *the external force*. We name this function and this type of model "the Newtonian dynamic model of an PO" (NDM ), its parameters "the NDM-kinetic constants", the corresponding trajectory "the NDM-trajectory", its velocity "the NDM-velocity", and its acceleration "the NDM-acceleration". For the small and slowly changing external force we suggest that without solving an equation of motion, *the long-time asymptotic dynamics* of an PO as $t \nearrow \infty$ (LTAD for short) can be adequately modeled by polynomials in time-derivatives of the external force, which we name "LTAD-models". By an LTAD-model we can take direct account of the different kinetic constants of an PO that supplement the PO-mass.

*In Sect.2, to collate some relevant facts about the possible* long-time asymptotic dynamics of an PO as $t \nearrow \infty$, we calculate two specific non-relativistic NDMs, which have three kinetic constants. In Sect.3, from these NDMs we calculate approximations of theirs LTADs by the polynomials in time-derivatives of the external force. These approximations provide a general understanding of the subject of this paper , i.e. the modeling of the LTAD by the polynomials in time-derivatives of the external force without solving an equation of motion. To illustrate this approach to modeling of the long-time asymptotic dynamics of an PO by LTAD-models, we make also the LTAD-models implied by the following five ordinary differential equations without solving them:

(i) the second order linear differential equation for a driven damped harmonic oscillator,

(ii) a system of two second order linear differential equation for two connected point-masses,

(iii) the Lorentz-Abraham-Dirac relativistic differential equation,

(iv) the Riccati differential equation of motion for the quadratic drag force, and

(v) a second order cubic nonlinear differential equation for a driven damped oscillator.

*We do not consider the theoretical physics problem of constructing an appropriate mathematical model for an electrified PO or for any other kind of PO.*

## 2. Generalizations of Newton's second law

2.1 *Driven damped harmonic oscillator*

Let us consider the NDM of an PO based on the point-mass with mass $m \geq 0$, which is moving along the *x*-axis under the influence of the external force $f(t), f(0)= 0$, and initially at rest at $x(0) = 0$. This point-mass is attached to the zero-length spring with the force constant $k \geq 0$, and slowed down by the frictional force with the non-negative viscous damping coefficient $c$ such that $c > 0$ if $km \neq 0$. Thus for all $t \geq 0$, the PO-position $x(t)$ satisfies the differential equation of motion for a driven damped harmonic oscillator:

$$m\, x^{(2)} + c\, x^{(1)} + k\, x = f \quad \text{with} \quad x^{(n)} \equiv (d/dt)^n\, x, \; n = 0, 1, 2 \ldots ; \qquad (1)$$

where *m, c,* and *k* are three non-negative NDM kinetic constants. Thus the NDM-trajectory

$$x(t) = \int_0^t z(t') f(t-t')\, dt' \quad \text{if} \quad m > 0, c > 0, \text{ and } k > 0, \qquad (2)$$

where the Green function

$$z(t) \equiv (m\sqrt{1-\zeta^2}\,\omega_0)^{-1} \exp(-\zeta\omega_0 t) \sin\sqrt{1-\zeta^2}\,\omega_0 t : \qquad (3)$$

$\omega_0 = \sqrt{k/m}$ is named "the un-damped angular frequency", and $\zeta = c/2m\omega_0$ is named "the damping ratio". The NDM-acceleration as a function of the external force $f(t)$ is given by:

$$x^{(2)}(t) = m^{-1} f(t) - m^{-1} \int_0^t z(t')[\, kf(t-t') + cf^{(1)}(t-t')]dt' \quad \text{if } m > 0, c > 0, k > 0; \qquad (4)$$

$$x^{(2)}(t) = m^{-1} f(t) - c\, m^{-2} \int_0^t \exp(-ct'/m)\, f(t-t')\, dt' \quad \text{if } m > 0, c \geq 0, k = 0; \qquad (5)$$

$$x^{(2)}(t) = c^{-1} f^{(1)}(t) - k\, c^{-2} \int_0^t \exp(-kt'/c)\, f^{(1)}(t-t')\, dt' \quad \text{if } m = 0, c > 0, k \geq 0; \qquad (6)$$

see also the equations (17), (18) and (19).

*Remarks*

If the external force $f(t)= 0$ for all $t \geq t_1$, then the differential equation of motion (1) implies

$$x(t) = a_1 \exp(-\zeta\omega_0 t) \sin(\sqrt{1-\zeta^2}\,\omega_0 t + \varphi) \quad \text{if } m > 0, c \geq 0, k > 0; \qquad (7)$$

$$x^{(1)}(t) = a_2 \exp(-ct/m) \quad \text{if } m > 0, c \geq 0, k = 0; \qquad (8)$$

$$x(t) = a_3 \exp(-kt/c) \quad \text{if } m = 0, c > 0, k > 0; \qquad (9)$$

where the four constants $a_1, a_2, a_3$, and $\varphi$ are determined by $f(t)$, $0 < t < t_1$. Thus, if $c > 0$, then (i) *after the cessation of the external force* $f(t)$ *the NDM-velocity* $x^{(1)}(t)$ *eventually stops changing,* and (ii) the properties of the external force $f(t)$ within any finite period of time have negligible effects on the NDM-velocity $x^{(1)}(t)$ as $t \nearrow \infty$ because this NDM is linear. However, were $c < 0$, then there would be the self-acceleration.

Modeling of the long-time asymptotic dynamics of a point-like object

## 2.2 *NDM of an PO based on two connected point-masses*

Let us consider two point-masses of equal mass $m \geq 0$, which are located on the x-axis, initially resting at points $x(0)= 0$ and $x_1(0)= 0$: thus $x^{(1)}(0)= 0$ and $x_1^{(1)}(0)= 0$. They are *connected* by the zero-length spring with the force constant $k/2 > 0$. The point-mass with the trajectory $x(t)$ is accelerated by the external force $f(t)$, $f(0)= 0$, and slowed down by the frictional force $-cx^{(1)}(t)$ with the viscous damping coefficient $c \geq 0$. Whereas the point-mass with the trajectory $x_1(t)$ is only slowed down by the frictional force $-cx_1^{(1)}(t)$. Thus the equations of motion for this system of two connected point-masses read:

$$mx^{(2)} = -cx^{(1)} + f - \tfrac{1}{2} k (x - x_1) \quad \text{and} \quad mx_1^{(2)} = \tfrac{1}{2} k (x - x_1) - cx_1^{(1)}. \tag{10}$$

So the velocity of the first point-mass is given by:

$$x^{(1)}(t) = \int_0^t z(t - t') [f^{(1)}(t') + k/(2m) \int_0^{t'} \exp(-c\tau/m) f(t' - \tau)d\tau]dt' \text{ if } m > 0, c \geq 0; \tag{11}$$

and

$$x^{(1)}(t) = c^{-1} \int_0^t \exp(-kt'/c) [k/(2c) f(t - t') + f^{(1)}(t - t')]dt' \text{ if } m = 0, c > 0. \tag{12}$$

For $c > 0$ we may consider $x(t)$ as an NDM-trajectory. And the time-differentiation of the equation (11) or (12) defines directly the corresponding NDM-acceleration.

*Remarks*

a) If the external force $f(t) = 0$ for all $t \geq t_1$, and $c = 0$, then the equations of motion (10) imply that after the cessation of the external force $f(t)$ the acceleration of the first point-mass is given by:

$$x^{(2)}(t) = a \sin(\sqrt{k/m}\, t + \varphi) \text{ for all } t \geq t_1, \tag{13}$$

where the real constants $a$ and $\varphi$ are determined by the external force $f(t)$, $0 < t < t_1$. So the velocity $x^{(1)}(t)$ of the first point-mass does not abide by Newton's first law: "The velocity of a body remains constant unless the body is acted upon by an external force." However, the center of mass $\tfrac{1}{2}(x + x_1)$ does abide by it, because its acceleration $\tfrac{1}{2}(x + x_1)^{(2)} = 0$ for all $t \geq t_1$, by (10) with $c = 0$.

b) According to the differential equations of motion (10), the differential equation of motion for the NDM-trajectory $x(t)$ is given by:

$$m^2 x^{(4)} + 2cm\, x^{(3)} + (km + c^2) x^{(2)} + kc\, x^{(1)} = \tfrac{1}{2} k f + c f^{(1)} + m f^{(2)}. \tag{14}$$

## 3. A small and slowly changing external force

Let us consider the LTAD in the case of the external force $\lambda F(\lambda t)$ whose magnitude and rate of change are determined by a small positive auxiliary parameter $\lambda$. First we approximate the LTADs of the NDMs (4) and (5) by polynomials in time-derivatives of the external force $\lambda F(\lambda t)$. Then we give a few exemplary modeling of the LTAD by polynomials in time-derivatives of the external force without solving an equation of motion.

### 3.1 *Approximations of a linear LTAD*

Let us consider the approximations of the linear LTADs implied by the NDMs (4) and (5) if the external force $f(t) = \lambda F(\lambda t)$ and $\lambda$ is small. Using the Taylor series expansion of the external force $\lambda F(\lambda t)$, we obtain from NDMs (4) and (5) the following two expansions in powers of $\lambda$:

$$x^{(2)}(t) = k^{-1}\lambda F^{(2)}(\lambda t) - k^{-2}c\lambda F^{(3)}(\lambda t) + O(\lambda^5) \quad \text{as } t \nearrow \infty \quad \text{if } m > 0,\ c > 0,\ k > 0 ; \qquad (15)$$

$$x^{(2)}(t) = c^{-1}\lambda F^{(1)}(\lambda t) - mc^{-2}\lambda F^{(2)}(\lambda t) + O(\lambda^4) \quad \text{as } t \nearrow \infty \quad \text{if } m > 0,\ c > 0,\ k = 0 ; \qquad (16)$$

provided $\sup_{t\geq 0}|F^{(n)}(\lambda t)| \leq \infty$ for $n = 0, 1, 2, 3$.

Whereas the differential equation of motion (1) implies that for all $t \geq 0$ and any $\lambda$:

$$x^{(2)}(t) = m^{-1}\lambda F(\lambda t) \quad \text{if } m > 0,\ c = 0,\ k = 0 ; \qquad (17)$$

$$x^{(1)}(t) = c^{-1}\lambda F(\lambda t) \quad \text{if } m = 0,\ c > 0,\ k = 0 ; \qquad (18)$$

$$x(t) = k^{-1}\lambda F(\lambda t) \quad \text{if } m = 0,\ c = 0,\ k > 0 . \qquad (19)$$

The equations (15)–(19) suggest that without having an equation of motion we may model the long-time asymptotic PO-trajectory, PO-velocity, or PO-acceleration by a sum of N time-derivatives of the small and slowly changing external force $\lambda F(\lambda t)$, say,

$$x^{(j)}(t) = \sum_1^N k_n \lambda F^{(n-1)}(\lambda t) + O(\lambda^{N+1}) \quad \text{where } j = 0, 1, 2 \qquad (20)$$

respectively, provided $\sup_{t\geq 0}|F^{(n)}(\lambda t)| \leq \infty$ for $n = 0, 1,..., N$. We name such a model of the LTAD for the small and slowly changing external force $\lambda F(\lambda t)$ "a LTAD-model", and the real constants $k_n$ "the long-time asymptotic kinetic constants". Note that by (20):

$$x^{(j+n)}(t) = O(\lambda^{n+1}) ,\quad n = 0, 1, 2, \ldots\ ,\ \text{as } t \nearrow \infty . \qquad (21)$$

If $k_n = 0$ for $n = 1, 2, \ldots, o-1$, and $k_o \neq 0$, $o \geq 1$, then the LTAD-model (20) implies the following novel *differential equation about the long-time asymptotic dynamics* of an PO:

$$\sum_j^N c_n x^{(n)}(t) = \lambda F^{(o-1)}(\lambda t) + O(\lambda^{o+N-1}) \quad \text{where } j = 1, 2, 3, \qquad (22)$$

which we name "the asymptotic differential equation", and vice versa. In particular, for $N = 4$, $j = 2$, and $k_1 \neq 0$ we have

$$c_2 = k_1^{-1},\ \ c_3 = -k_1^{-2}k_2,\ \text{and}\ \ c_4 = k_1^{-3}k_2^2 - k_1^{-2}k_3 ; \qquad (23)$$

Modeling of the long-time asymptotic dynamics of a point-like object

and
$$k_1 = c_2^{-1}, \quad k_2 = -c_3 c_2^{-2}, \quad \text{and} \quad k_3 = c_2^{-3} c_3^2 - c_2^{-2} c_4. \tag{24}$$

*Remarks*

a) The start-up dynamics of the NDM (4) is given by:
$$x^{(2)}(t) = m^{-1}\lambda^2 F^{(1)}(0)\, t + O(\lambda^3 t^2). \tag{25}$$
But the way how the long-time asymptotic dynamics of the NDM (4), the LTAD-model (15) depends on the NDM kinetic constants m, c, and k is fundamentally different.

b) By definition a LTAD-model contains *only one* time-derivative of the NDM-trajectory; an asymptotic differential equation contains by definition *only one* time-derivative of the external force ; while the equation of motion for the NDM-trajectory is not so constrained, but it has *only a finite* number of terms.

c) The estimate (21) with j = 0 implies the (21) with j =1 and j = 2, but not vice versa!

d) The equations (4) and (15) show that the NDM and the corresponding LTAD-model may differ significantly. But if the external force is small and slowly changing, then we may always use the LTAD-models to calculate approximations of the long-term asymptotic NDM-trajectories.

e) In contrast to the differential equation of motion (1) that depends continuously on the kinetic constants $m$, $c$, and $k$, the corresponding LTAD-models may not in view of (15) and (16). The same goes for the corresponding asymptotic differential equations (22), which in general do not determine the differential equation of motion (1)

f) Eliminating the higher time-derivatives of the trajectory by iteration from an NDM differential equation of motion like (1) or (14), we can calculate directly the corresponding LTAD-models without solving an equation of motion, provided they exist! So assuming that the estimate (21) is correct, we could calculate from the NDM differetial equation of motion (14) for two connected point-masses the correspondig LTAD-model:
$$x^{(1)}(t) = (2c)^{-1} \lambda\, F(\lambda t) + O(\lambda^2) \quad \text{if } m \geq 0,\ c > 0, \tag{26}$$
and
$$x^{(2)}(t) = (2m)^{-1} \lambda\, F(\lambda t) + O(\lambda^3) \quad \text{if } m > 0,\ c = 0. \tag{27}$$
However, since (21) is false if c = 0, the equation (27) is false, cf. the equation (13).

## 3.2. Relativistic LTAD-models

Let us consider a relativistic NDM based on the point-mass with mass $m \geq 0$, which is located at $\mathbf{r}(t)$ and moving with velocity $\mathbf{v}(t)$ under the influence of the external force $\lambda \mathbf{F}(\lambda t)$ with the small auxiliary parameter $\lambda > 0$. We define the external four-force

$$\lambda \Phi(\lambda t) \equiv \gamma\big(\boldsymbol{\beta}(t) \cdot \lambda \mathbf{F}(\lambda t),\ \lambda \mathbf{F}(\lambda t)\big), \text{ where } \gamma(t) \equiv 1/\sqrt{1 - |\boldsymbol{\beta}|^2} \text{ with } \boldsymbol{\beta}(t) \equiv \mathbf{v}/c, \quad (28)$$

and the NDM four-velocity $\beta(t) \equiv (\gamma, \gamma\boldsymbol{\beta})$: we use the metric with the signature $(+ - - -)$, so $\beta \cdot \beta = 1$. We introduce an additional four-force $\Delta(t)$, which depends on the external force $\lambda \Phi(\lambda t)$, and formulate a relativistic NDM as follows:

$$mc\beta^{[1]}(t) = \Delta(t) + \lambda \Phi(\lambda t) \quad \text{with} \quad \beta^{[n]} \equiv (\gamma d/dt)^n \beta, \quad n = 0, 1, 2, \ldots, \quad (29)$$

where $t/\gamma$ is the proper time. As $\beta \cdot \beta^{[1]} = 0$ and $\beta \cdot \Phi = 0$, we may rewrite the relativistic NDM (29) as

$$mc\beta^{[1]}(t) = (1 - \beta\,\beta\,\cdot)\Delta(t) + \lambda \Phi(\lambda t). \quad (30)$$

Generalizing the linear LTAD-model (20) with $j=2$, we model the four-force $\Delta(t)$ by a polynomial in time-derivatives $\lambda \Phi^{[n]}$ so as to get a relativistic LTAD-model:

$$\begin{aligned}\beta^{[1]} = (1 - \beta\,\beta\,\cdot)\big[&k_1\,\lambda\Phi + k_2\,\lambda\Phi^{[1]} + k_{31}\lambda^3(\Phi \cdot \Phi)\Phi + k_{32}\,\lambda\Phi^{[2]} + k_{41}\,\lambda^3(\Phi^{[1]} \cdot \Phi)\Phi \\ &+ k_{42}\,\lambda^3(\Phi \cdot \Phi)\Phi^{[1]} + k_{43}\,\lambda\Phi^{[3]} + k_{51}\lambda^5(\Phi \cdot \Phi)^2 \Phi + k_{52}\,\lambda^3(\Phi^{[1]} \cdot \Phi^{[1]})\Phi \\ &+ k_{53}\,\lambda^3(\Phi^{[1]} \cdot \Phi)\Phi^{[1]} + k_{54}\,\lambda^3(\Phi^{[2]} \cdot \Phi)\Phi + k_{55}\,\lambda^3(\Phi \cdot \Phi)\Phi^{[2]} + k_{56}\,\lambda\Phi^{[4]}\big] + O(\lambda^6),\end{aligned} \quad (31)$$

where the real constants $k_1, \ldots, k_{56}$ are independent of the external four-force $\lambda \Phi(\lambda t)$. We name them "the relativistic long-time asymptotic kinetic constants", as they specify a relativistic LTAD up to the order of $\lambda^5$ inclusive in the case of a small and slowly changing external four-force $\lambda \Phi(\lambda t)$. We see no physical reason to believe that the number of the independent relativistic long-time asymptotic kinetic constants might be small.

In general, when modeling of the LTAD for a given PO, we expect the equation (31) to be an appropriate relativistic LTAD-model provided that

a) each of its term is essentially smaller than the preceding one,
b) the properties of the PO are permanent,
c) the values of the external four-force $\lambda \Phi(\lambda t)$ within any finite period of time have negligible effects on the LTAD,
d) the PO-acceleration depends without delay on the external four-force $\lambda \Phi(\lambda t)$.

The NDMs (17)–(19) suggest that taking only a finite number of terms in the equation (31), we might actually get an adequate relativistic NDM equation of motion, which generalizes Newton's second law, exhibiting no self-acceleration with runaway solutions

If $\Delta(t) = \Delta(-t)$, then the NDM (30) is invariant under time reversal and the potential relativistic LTAD-model is given by the equation (31) with $k_2 = k_{41} = k_{42} = k_{43} = 0$.

Eliminating by iterations all the time-derivatives $\lambda \Phi^{[n]}$ except one of $\lambda \Phi^{[0]}$ from the relativistic LTAD-model (31), we get a relativistic asymptotic differential equation, analogous to the asymptotic differential equation (22).

### 3.3 *Relativistic LTAD-models of an electrified PO*

In 1673, Huygens derived the formula for the period of a mathematical pendulum. However, if we rub a real pendulum and electrify it, the classical electrodynamics has provided so far no adequate dynamic model of such an electrified PO. In 1892, H. A. Lorentz started an ongoing quest to take account of the radiation reaction force (the effect of the loss of four-momentum by the electromagnetic radiation) in modeling of the motion of a classical charged particle, cf. [1]. So there is a century old open question how to construct within the framework of the classical electrodynamics an adequate dynamic model of an electrified PO that has at least two kinetic constants, mass and charge, and no unsatisfactory features such as the runaway solutions. For a survey of many attempts to solve this problem and a physically consistent solution proposed by Ford–O'Connell in 1991 see O'Connell [2], and the references cited therein.

To show how we can use *a relativistic asymptotic differential equation to take account of the assumptions about the dynamic properties of PO whose LTAD we are modeling*, we now collate the basic *relativistic* premises of Dirac's theory of an electrified PO. Presuming that this PO is electrified by a point-like charge, we follow Schott [3] and express the additional four-force as

$$\Delta = - d(\beta^{[1]} \cdot \beta^{[1]})\beta + B^{[1]}, \qquad (32)$$

i.e. as the difference between the intensity $d(\beta^{[1]} \cdot \beta^{[1]})\beta$, $d \geq 0$, of the four-momentum emitted by the Liénard-Wiechert potentials with the cyclically moving singularity at $r(t)$, cf. [1, §6.6], and the time-derivative of an "acceleration four-momentum $B(t)$". So for an electrified PO we rewrite the relativistic NDM (29) as follows:

$$mc\beta^{[1]} - d(\beta^{[1]} \cdot \beta^{[1]})\beta + B^{[1]} = \lambda \Phi. \qquad (33)$$

Assuming that the acceleration four-momentum $B(t)$ is a four-function of the time-derivatives $\beta^{[n]}$ and $\Phi^{[n]}$, Dirac [4] concluded that the conservation of four-momentum requires that

Modeling of the long-time asymptotic dynamics of a point-like object

$$\beta \cdot (B + d\beta^{[1]})^{[1]} = 0 \ . \tag{34}$$

Thereafter, Bhabha [5] pointed out that the conservation of angular four-momentum requires that the cross product

$$\beta \wedge (B + d\beta^{[1]}) \tag{35}$$

is a total differential with respect to the proper time.

In [1, Sects.10.1 and 10.2] and [6], we pointed out seventeen qualitative properties that we are expecting from a physically realistic NDM for an electrified PO. So far no such NDM is known.

Inspired by the asymptotic differential equation (22), we assume that the nth time-derivative $\beta^{[n]}$ is of the order $\lambda^n$ as $t \nearrow \infty$, and approximate the time-derivative $B^{[1]}$ in the NDM (33) by the polynomials in $\beta^{[n]}$, cf. [1, Ch.9] and [7]. Accordingly, for an electrified PO the relativistic asymptotic differential equation about the long-time asymptotic acceleration is given up to the order of $\lambda^2$ inclusive by:

$$mc\beta^{[1]} - d(1 - \beta\,\beta\,\cdot)\beta^{[2]} = \lambda\Phi \ . \tag{36}$$

In [8] we gave such a relativistic asymptotic differential equation for an electrified PO up to the order of $\lambda^6$ inclusive. The equation (36) is traditionally not taken as a relativistic asymptotic differential equation, but as the Lorentz-Abraham-Dirac *equation of motion*, though it exhibits the self-acceleration. The first term is due to Einstein and the second one is due to Dirac [4], who assumed that an electron is such a simple thing that the equation (36) ought to be the correct equation of motion with $d = e^2/6\pi\epsilon_0 c^2$. So we name the constant $d$ "the Dirac kinetic constant".

Eliminating by iteration the time-derivative $\beta^{[2]}$ from the asymptotic differential equation (36), we get for an electrified PO the following LTAD-model:

$$mc\beta^{[1]} = (1 - \beta\,\beta\,\cdot)\bigl[\lambda\Phi + \lambda d/mc\,\Phi^{[1]}\bigr] + O(\lambda^3) , \tag{37}$$

analogous to the LTAD-model (31). In [1, Sect.11.4] we gave for an electrified PO such a relativistic LTAD-model up to the order of $\lambda^6$ inclusive. According to the LTAD-model (37), in the case of a small and slowly changing external four-force $\lambda\Phi(\lambda t)$, the mass $m$ and next the Dirac kinetic constant $d$ are the most important *relativistic long-time asymptotic kinetic constants of any electrified PO.*

3.4 *The quadratic drag force*

To model the dynamics of an PO moving through a fluid at relatively large velocity, we take the point-mass with mass m ≥ 0, which is moving along the x-axis under the influence of the external force $\lambda F(\lambda t) > 0$. This PO is slowed down by the quadratic drag force $c_d(x^{(1)}(t))^2$,

Modeling of the long-time asymptotic dynamics of a point-like object

$c_d > 0$. Thus the PO-velocity $x^{(1)}$ satisfies the following Riccati differential equation:

$$m\,x^{(2)} + c_d(x^{(1)})^2 = \lambda F(\lambda t). \tag{38}$$

If the estimates

$$x^{(n)}(t) = O(\lambda^{n-1/2}),\ n = 1, 2, \dots,\ \text{as } t \nearrow \infty, \tag{39}$$

are true also for $m > 0$, the differential equation (38) implies the new type of LTAD-models:

$$x^{(1)}(t) = \sqrt{\lambda/c_d}\,\sqrt{F(\lambda t)} - \tfrac{1}{4}m\,\sqrt{\lambda/c_d}\,F^{(1)}(\lambda t)/\sqrt{F(\lambda t)} + O(\lambda^{5/2})\ \text{as } t \nearrow \infty. \tag{40}$$

We can improve this model by adding the frictional force $-c x^{(1)}(t)$ to the equation (38).

3.5 *The strong string*

By using a strong string, we generalize the differential equation of motion (1) for a driven damped harmonic oscillator so that

$$m\,x^{(2)} + c\,x^{(1)} + k_1\,x + k_3 x^3 = \lambda F(\lambda t), \tag{41}$$

where $m, c, k_1, k_3 \geq 0$ and $c > 0$ if $(k_1 + k_3)m > 0$. We rewrite this differential equation, by Cardano's formula as follows

$$x = \left[-q + (q^2 + p^3)^{1/2}\right]^{1/3} - \left[q + (q^2 + p^3)^{1/2}\right]^{1/3}, \tag{42}$$

where $p \equiv k_1/3k_3$, $q \equiv (r - \lambda F(\lambda t))/2k_3$ and $r \equiv m\,x^{(2)} + c\,x^{(1)}$. Presuming that the estimates

$$x^{(n)}(t) = O(\lambda^{n+2\mathrm{sig}(k_1)/3 + 1/3}),\ n = 0, 1, \dots,\ \text{as } t \nearrow \infty, \tag{43}$$

are true also if $m, c > 0$; using Taylor series of $x \equiv [\mathrm{rhs}(42)]$ as a function of $r$; and eliminating the time-derivatives $x^{(n)}$, we infer the following LTAD-model:

$$x = x^{\{0\}} + c\,x^{\{1\}}(x^{\{0\}})^{(1)} + O(\lambda^{7/3+2\mathrm{sig}(k_1)/3})\ \text{as } t \nearrow \infty,\ \text{with } x^{\{n\}} \equiv (\partial/\partial r)^n[\mathrm{rhs}(42)]\ \text{at } r = 0. \tag{44}$$

4. Conclusions

We introduce four original concepts:

a) First, the point-like object (PO) specified as a classical extended real object whose response to an external force is aptly specified solely by the trajectory of a single point, whose velocity eventually stops changing after the cessation of the external force.

b) Second, the dynamic model of an PO (NDM) that generalizes Newton's second law by the explicit modeling of PO-acceleration by a nonlinear function of the external force.

c) Third, the long-time asymptotic dynamics of an PO (LTAD) modeled by polynomials in time-derivatives of the external force (by LTAD-models), e.g. for the cyclic motion.

d) Fourth, the asymptotic differential equations about the LTAD.

Modeling of the long-time asymptotic dynamics of a point-like object

Given a differential equation of motion for an PO, we can calculate the corresponding LTAD-models directly to any order without solving this differential equation of motion! However, these LTAD-models generally do not specify the original equation of motion.

An LTAD-model implies a novel, asymptotic differential equation about the LTAD, and vice versa! We can use an LTAD-model for calculating the approximations of the long-time asymptotic PO-trajectories as $t \nearrow \infty$; whereas in general, we may not use an asymptotic differential equation to this end, but we can use it for taking account of the dynamic properties, e.g. at the cyclic motion, of an PO whose LTAD-models we are making. We may use either to access those long-time asymptotic kinetic constants that supplement the PO-mass, e.g. a charge, by observing the long-time asymptotic PO-trajectories, cf. [8].

Though there are infinitely many possible NDMs for a particular PO, there are only a very few theoretical principles how to construct one. Nevertheless, using an appropriate, generic ansatz such as (31), (40) or (44) for an LTAD-model, enables us to specifically model the data about the long-time asymptotic trajectories of a particular PO without knowing an adequate NDM.